\documentclass[12pt]{JHEP3}
\usepackage{amsmath,amsfonts,amssymb,amsthm}
\usepackage{epsf}

\newtheorem{thm}{Theorem}[section]

\newtheorem{prop}[thm]{Proposition}
\theoremstyle{definition}
\newtheorem{defn}[thm]{Definition}
\theoremstyle{remark}

\numberwithin{equation}{section}

\author{Boyu Hou, Sen Hu, Yanhong Yang\\
$^1$ Department of physics, Northwest University, Xi'an, Shannxi, \\
$^2$ Department of Mathematics, University of Science and Technology of China, Hefei, Anhui, 230026\\
$^\dagger$ \email{shu@ustc.edu.cn}}

\title{On special geometry of the moduli space of string vacua with fluxes}

\abstract{In this paper we construct a special geometry over the
moduli space of type II string vacua with both NS and RR fluxes
turning on. Depending on what fluxes are turning on we divide into
three cases of moduli space of generalized structures. They are
respectively generalized Calabi-Yau structures, generalized
Calabi-Yau metric structures and ${\cal N} =1$ generalized string
vacua. It is found that the $d d^{\cal J}$ lemma can be established
for all three cases. With the help of the $d d^{\cal J}$ lemma we
identify the moduli space locally as a subspace of $d_{H}$
cohomologies. This leads naturally to the special geometry of the
moduli space. It has a flat symplectic structure and a K$\ddot{\rm
a}$hler metric with the Hitchin functional (modified if RR fluxes
are included) the K$\ddot{\rm a}$hler potential. Our work is based
on previous works of Hitchin and recent works of Gra$\tilde{\rm
n}$a-Louis-Waldram, Goto, Gualtieri, Yi Li and Tomasiello. The
special geometry is useful in flux compactifications of type II
string theories.}

\keywords{Generalized structure, string vacua, special geometry}
\preprint{}

\begin{document}

\section{\textbf{Introduction}}

String compactifications of flux backgrounds are of great interests
in recent years. It is hoped that in the general background more
realistic models can be constructed, moduli stability problem of
inflation models can be tracked and more general black hole
attractors can be obtained, among other things. There have been much
developments in this direction. For recent reviews see \cite{Grana},
\cite{DK}.

The notion of generalized Calabi-Yau manifolds was introduced by
Hitchin in \cite{Hitchin} and was further studied by Gualtieri
\cite{Gua}. Hitchin \cite{Hitchin} studies special geometry of the
moduli space of generalized complex structures. Their work attracts
great interests of string theorists and generated much works in the
past years.

In string compactifications a crucial step is to establish special
geometry of the moduli space of generalized Calabi-Yau manifolds.
Special geometry of generalized Calabi-Yau manifolds which
incorporate all fluxes was proposed and established by
Gra$\tilde{\rm n}$a-Louis-Waldram in \cite{GLW1}, \cite{GLW2}. In
this paper we shall give an alternative treatment of the moduli
space of string vacua of generalized structures which will lead
naturally to such a special geometry.

Depending on what fluxes are turning on we divide into three cases
of moduli space of generalized structures. They are respectively
generalized Calabi-Yau structures, generalized Calabi-Yau metric
structures and ${\cal N} =1$ generalized string vacua.

To incorporate Neveu-Schwarz fluxes we change exterior derivative
$d$ to $d + H \wedge$. To include RR fluxes we introduce local
sources, as anticipated from the no-go theorem. We shall work on a
non-compact manifolds with divisors consisting of generalized
complex submanifolds (with respect to the integrable generalized
complex structure) and orientifolds as local sources.

The moduli space of generalized structures can be defined as the
space of generalized structures over a manifold modulo
diffeomorphisms and exact one forms. Deformations of generalized
manifolds were studied by Li Yi \cite{LiYi} and Goto \cite{Goto}.
They construct a complex of Lie algebroid and prove
unobstructed-ness for the moduli space of generalized structures
over a manifold. We borrow a crucial argument from Goto that the
period map is injective whenever the $\partial \bar{\partial}$ lemma
is true. Such a lemma is provided by Gualtieri for twisted
generalized K$\ddot{\rm a}$hler manifolds. We extend the $\partial
\bar{\partial}$ lemma for all ${\cal N} = 1$ generalized vacua. This
would help us to identify the moduli space as co-chains represented
by pure spinors.

As long as we have a good identification of the moduli space it
leads naturally to the special geometry of the moduli space. We
shall prove that there is a natural symplectic form, a complex
structure and a K$\ddot{\rm a}$hler metric with logarithm of Hitchin
functional as its potential function. We modify the Hitchin
functional so that RR fluxes arise as source terms. The critical
points of the new Hitchin functional are precisely the
super-symmetric solutions of type II strings with all fluxes turning
on. Those are the basis of special geometry for the moduli space of
super-symmetric solutions.

\section{Geometry of $G$ structures}

\subsection{Moving frames: $G$ structures}

 In the study of geometry it is important to find out interesting
 geometric structures to work with. It turns out that
 $G$ structures and subsequently generalized $G$ structures are of
 great interests from both geometric and physical points of view.

 Given a manifold $M$ we consider the frame bundle of the tangent space.
The frame bundle is a bundle with fiber $Gl(n, {\bf R})$. A $G$
structure is a subbundle of the frame bundle whose fiber is $G$. We
also say that we have the structure group reduction to $G$.

 For example, if we are given a metric $g$ then we may consider all
orthonormal frames. This gives a subbundle whose fiber is $O(n, {\bf
R})$. If the maifold is orientable we would have the structure group
reduction to $Sl(n, {\bf R})$.

 If we are given an almost complex structure, i.e. $J: TM \rightarrow TM, J^{2} =
 -Id$, then we have  $T \otimes {\bf C} = T^{1,0} \oplus T^{0,1}$. This way we have
 the structure group reduction $Gl(n, {\bf C}) \subset Gl(2n, {\bf
 R})$.

 The existence of an $G$ structure is a topological problem. It turns out
 that the interesting geometric objects are those $G$ structures which are integrable.
 Here is a definition of integrability of a $G$ structure \cite{Ch}:

 A $G$ structure is called holonomic or integrable if: Under the Levi-Civita
 parallelism permissible frames remains permissible.

 For example, an almost complex structure is integrable iff $\nabla J = 0$.

 An integrable $G$ structure is a manifold with special holonomy
 with the holonomy group $G$. Those manifolds of special holonomy
 has been classified by M. Berger into seven classes provided that
 they are not homogeneous manifolds.

 1) $G = SO(n)$, Orientable Riemannian manifolds;

 2) $G = U(n)$, K$\ddot{\rm a}$hler manifolds.

 There are five classes of manifolds of special holonomy on which
 there exists a non-zero spinor $\epsilon$ such that $\nabla
 \epsilon  = 0$:

 3) $G = SU(n)$, Calabi-Yau manifolds;

 4) $G = Sp(n)$, hyper-K$\ddot{\rm a}$hler manifolds;

 5) $G= Sp(n) \times Sp(1)$, Quaternionic K$\ddot{\rm a}$hler manifolds;

 6) $G = G_{2}$, seven dimensional manifolds;

 7) $G = Spin(7)$, eight dimensional manifolds.

  One may relax the integability condition of a Levi-Civita connection to a
connection with non-vanishing torsion. This gives Chern connection
and Bismut connection. A connection $\nabla + T$ is compatible with
the metric if and only if $T$ is an anti-symmetric tensor. It turns
out that the torsion gives NS fluxes in string vacua. In recent
years several authors got new solutions of string vacua with NS
fluxes. See \cite{GMW, MS} and references in for more details.

\subsection{Special geometry over the moduli space of Calabi-Yau manifolds}

A complex manifold of $SU(3)$ holonomy is called a Calabi-Yau
manifold. It appears in the compactification of string theory. The
key property of a Calabi-Yau manifold is the existence of a
non-vanishing spinor which is parallel with respect to the spin
connection \cite{BBS}:

$$\nabla_{\mu} \epsilon = 0.$$

By Yau's theorem, any K$\ddot{\rm a}$hler manifold $M$ with
$c_{1}(TM) = 0$ is a Calabi-Yau manifold. For such a manifold, there
exists a unique Ricci flat metric with a given (1,1) closed form
$\omega$, such that $\int_{M} \omega \wedge \omega \wedge \omega >
0$.

The Calabi-Yau structure (the above Ricci flat metric) depends on a
symplectic structure $\omega$ and a complex structure and the moduli
space of Calabi-Yau structures is:

${\cal M} = {\cal M}_{K} \times {\cal M}_{C} \subset H^{1,1}(M)
\times H^{2,1}(M).$

It turns out that the moduli space itself is a K$\ddot{\rm a}$hler
manifold with the K$\ddot{\rm a}$hler metric. We call this the
special geometry of the moduli space of Calabi-Yau manifolds. The
metric is:

$$ds^{2} = \frac{1}{V} \int_{M^{6}} g^{a \bar{b}} g^{c \bar{d}}
(\delta g_{ac} \delta g_{\bar{b} \bar{d}} + (\delta g_{a\bar{d}}
\delta g_{c \bar{b}} - \delta B_{a\bar{d}} \delta B_{c \bar{b}})
d^{6}x.$$

It is a K$\ddot{\rm a}$hler metric with K$\ddot{\rm a}$hler
potential:

$$e^{K^{2,1}} = - i \int \Omega \wedge \bar{\Omega},$$

$$e^{K^{1,1}} = - i \int \omega \wedge \omega \wedge \omega.$$

Much of developments in string theory depend on this special
geometry. For example one may use special geometry to study
variation of Hodge structures of Calabi-Yau manifolds. This gives a
topological B model. One may study black hole attractors. And of
course it is important to study string compactifications through the
special geometry. And one may study mirror symmetry. The problem is
given a Calabi-Yau manifold $M_{1}$ find a mirror Calabi-Yau
manifold $M_{2}$, such that

$${\cal M}_{K}(M_{1}) = {\cal M}_{C}(M_{2}).$$

\section{Generalized structures and string vacua with fluxes}

In the study of string theory we want to know string vacua with
fluxes turning on. It turns out the the setting of generalized
geometry is a good framework to construct new string vacua. The
purpose of this paper is to establish a mathematical foundation of
special geometry for such new string vacua. We start with some
elementary definitions.

\subsection{Spinors and bispinors}

Let $(V, q)$ be a vector space with a quadratic form $q$, then we
have a Clifford algebra:

\begin{equation}
Cl(V, q) = \oplus_{r \ge 0} V^{\otimes r} / \{uv + vu = q(u, v)
Id\}.
\end{equation}

It is known that, depending on whether the dimension of $V$ is odd
or even, there are one or two irreducible representations of $Cl(V,
q)$:

\begin{equation}
\rho: Cl(V, q) \rightarrow End(\Delta).
\end{equation}

Here $\Delta$ is the space of spinors. Now we consider a special
case. Let $T$ be a vector space and let $V = T \oplus T^{*}$, where
$T^{*}$ is the dual space of $T$. There is a natural pairing between
$T$ and $T^{*}$. So we have a quadratic form:

\begin{equation}
<X + \xi, Y + \eta > = \frac{1}{2}(\xi(Y) + \eta(X)).
\end{equation}

Now consider the Clifford algebra $Cl(V, <,>)$. We find that the
exterior algebra of $T^{*}$ is the natural space of spinors for
this Clifford algebra, i.e. if we consider

\begin{equation}
(X + \xi). \omega = i_{X} \omega + \xi \wedge \omega,
\end{equation}

\noindent we would have a natural representation:

\begin{equation}
\rho: Cl(V, <,>) \rightarrow End(\Lambda^{*}T^{*}).
\end{equation}

Given any $(T, g)$, where $g$ is a quadratic form over $T$, we have
a Clifford algebra. Let $\gamma^{\mu}$ be a basis of the Clifford
algebra. It satisfies:

\begin{equation}\{\gamma^{\mu}, \gamma^{\nu}\} = 2 g^{\mu \nu} Id.
\end{equation}

The representation of $Cl(T,g)$ gives rise to $(T,g)$ spinors. We
need to imbed $(T,g)$ into $(T \oplus T^{*})$ isometrically. This
can be done as follows:

\begin{equation}
X \rightarrow X + i_{X}g \pm i_{X}b,
\end{equation}

\noindent where $b$ is any anti-symmetric two form.  Let us
consider anti-symmetrized $\gamma$-matrices,

\begin{equation}
\gamma^{\mu_{1} ... \mu_{p}} = \gamma^{[\mu_{1}} ...
\gamma^{\mu_{p}]}.
\end{equation}

The precise relationship of $(T,g)$ spinors and $(T \oplus T^{*},
<,>)$ spinors is given by the so-called bi-spinor construction. Let
$\epsilon_{1}, \epsilon_{2}$ be two $(T,g)$ spinors, then

\begin{equation}\Sigma_{\mu_{1} < \mu_{2} < ... < \mu_{p}}
\bar{\epsilon}_{2} \gamma^{\mu_{1} ... \mu_{p}} \epsilon_{1}
e_{\mu_{1}} \wedge ... \wedge e_{\mu_{p}}
\end{equation}

\noindent is an element of $\Lambda^{*}(T^{*})$, here
$\{e_{\mu}\}_{\mu = 1,...,n}$ is a basis of $T^{*}$.

This gives rise to a map which is actually an isomorphism, we call
it the bispinor construction:

\begin{equation}
\Delta \otimes \Delta \cong \Lambda^{*} T^{*},\\
(\epsilon_{1}, \epsilon_{2}) \rightarrow \Sigma_{0 \le \mu_{1} <
\mu_{2} < ... < \mu_{p} \le n} \bar{\epsilon}_{2} \gamma^{\mu_{1}
... \mu_{p}} \epsilon_{1} e_{\mu_{1}} \wedge ... \wedge e_{\mu_{p}}.
\end{equation}

\subsection{Almost generalized complex structures}

\textbf{Definition:} An almost generalized complex structure over a
vector space $T \oplus T^{*}$ is an endomorphism:

\begin{equation}\mathcal{J}: T \oplus T^{*} \rightarrow T \oplus T^{*}, \quad \quad
 \hbox{ where }\quad  \mathcal{J}^{2} = -
Id, \quad \quad <\mathcal{J}., \mathcal{J}.> = <.,.>.\end{equation}

We have then the maximal isotropic space:

\begin{equation}L = \{ \mathcal{J} = i\} \subset (T \oplus T^{*})_{\otimes {\mathbb
 C}},\end{equation}

such that

\begin{equation}(T \oplus T^{*})_{\otimes {\mathbb C}} = L \oplus \bar{L},\quad\quad
 L \cap \bar{L} = \{0 \}.\end{equation}

We also call such a structure an almost Dirac structure. Recall
that $\Lambda^{*}T^{*}$ is the space of spinors for $(T \oplus
T^{*}, <,>)$, via the map:

\begin{equation}(X + \xi) . \omega = i_{X} \omega + \xi \wedge \omega.\end{equation}

$\Lambda^{ev/od}T^{*}$ of even and odd degrees are two irreducible
representations of the spin group. If there is a $\rho \in
\Lambda^{*}T^{*}$, such that

\begin{equation}L = \{X + \xi |(X + \xi). \rho = 0\}
\end{equation}

\noindent we call $\rho$ a pure spinor. Given a pure spinor $\rho$
from Proposition 2.16 we have that $\rho + i \hat{\rho} = 2
\varphi$ is a stable spinor. The condition that $L \cap \bar{L} =
\{ 0 \}$ is precisely the condition that $(\varphi, \bar{\varphi})
\ne 0$. We see that a pure spinor, modulo multiplication of a
nonzero constant, determines a maximal isotropic space of an
almost generalized complex structure. Each maximal isotropic
subbundle $L$ corresponds to a sub line bundle of pure spinors
$U$.

We can parameterize the space of generalized $G$ structures by using
stable spinors. For us the most important cases are six and seven
dimensional. The moduli space of almost generalized $SU(3) \times
SU(3)$ structures over a vector space is:

\begin{equation}\mathcal{M} = SO(6, 6) / SU(3, 3) = U_{\rho}/{\mathbb C}^{*},\end{equation}

\noindent where $U_{\rho}$ is the space of stable spinors. The
moduli space of almost generalized $G_{2} \times G_{2}$ structures
over a vector space is:

\begin{equation}\mathcal{M} = SO(7, 7) / G_{2} \times G_{2} = U_{\rho}/{\mathbb C}^{*},\end{equation}

\noindent where $U_{\rho}$ is the space of stable spinors.

To pass the definition of an almost generalized $G$ structure over a
vector space to a manifold we need the notion of bundles. Let
$M^{6}$ be a manifold, over the manifold we have a bundle:

\begin{eqnarray}
U_{\rho}/{\mathbb C}^{*} \rightarrow \mathcal{E} \nonumber \\
\downarrow \nonumber \\ M^{6}
\end{eqnarray}

The space of almost generalized $G$ structures is the space of
sections of the above bundle.

\textbf{Remark: }It is purely a topological condition whether an
almost generalized $G$ structure exists. A necessary condition is
the existence of a stable spinor. This condition is also a
sufficient condition for dimensions 6 and 7.

In general we define geometric structures as an orbit of ${\cal
B}(V)$ structures. Let $V$ be an n-dimensional vector space with a
dual space $V^{*}$. The conformal group $Cpin(V \oplus V^{*})$ of $V
\oplus V^{*}$ acts on direct sums of exterior algebra $\oplus^{l}
\Lambda^{*}V^{*}$. Let $\Phi = (\phi_{1}, ..., \phi_{l})$ be an
element of the direct sum $\oplus^{l} \Lambda^{*}V^{*}$ and ${\cal
B}(V)$ the orbit of $Cpin(V \oplus V^{*})$ through $\Phi$. We fix
the orbit ${\cal B}(V)$ and goes to a manifold $M^{n}$. The orbit
${\cal B}(V)$ yields the orbit in $\oplus^{l} \Lambda^{*}TM^{*}_{x}$
for each point $x \in M^{n}$. We then have a fibre bundle ${\cal
B}(M)$ by

$${\cal B}(M) := \cup_{x \in M^{n}} {\cal B}(T_{x}M) \rightarrow
M^{n}.$$

The set of global sections of ${\cal B}(M)$ is denoted by ${\cal
E}_{\cal B}(M)$. We define a ${\cal B}(V)$-structure by a
$d_{H}$-closed section of ${\cal E}_{\cal B}(M)$. In the next
section we shall explain that the $d_{H}$-closed condition is
precisely the integrability condition of the underlying almost
structure.

\subsection{Integrability of an almost generalized structure}

Given an almost generalized complex structure whether it is
integrable or not is of great interests. We will see that
integrable generalized $G$ structures are closely related to Type
II string vacua.

\textbf{Definition: (Integrability of an almost generalized complex
structure)} Let $\mathcal{J}$ be an almost generalized complex
structure and $L$ the maximal isotropic space associated to
$\mathcal{J}$. We say that $\mathcal{J}$ is integrable if $L$ is
Courant involutive, i.e. $L$ is closed with respect to the Courant
bracket:

\begin{equation}[X + \xi, Y + \eta] = [X, Y] + \mathcal{L}_{X} \eta - \mathcal{L}_{Y}
\xi -\frac{1}{2} d (i_{X} \eta - i_{Y} \xi).\end{equation}

This condition can be expressed in terms of pure spinors.

The Clifford algebra $CL(T \oplus T^{*})$ is filtered as follows:

\begin{eqnarray}
\mathbb{R}&=&CL^{0} < CL^{2} < ... < CL^{2n}=CL^{+}(T \oplus T^{*})\\
T \oplus T^{*}&=&CL^{1} < CL^{3} < ... < CL^{2n-1}=CL^{-}(T \oplus
T^{*}),
\end{eqnarray}

\noindent where $CL^{l}$ is spanned by products of numbers of not
more than $l$ elements of $T \oplus T^{*}$. The Clifford
multiplication respects this graded filtration structure.

Suppose we have a trivialization of the pure sub line bundle $U$
with a nonzero section $\rho$. We may decompose the space of
differential forms by Clifford multiplication on $\rho$:

\begin{eqnarray}
U &=& U_{0} < U_{2} < ... < U_{2n} = \Lambda^{ev/od} T^{*} \otimes
\mathbb{C}, \\
L^{*} .U &=& U_{1} < U_{3} < ... < U_{2n-1} = \Lambda^{od/ev} T^{*}
\otimes \mathbb{C},
\end{eqnarray}

\noindent where $U_{k}$ is defined as $CL^{k} . U, k = 1, ..., 2n,
U_{k}$ are eigenspaces of eigenvalue $-ik$ of ${\cal J}$ acting on
forms through spin representation.


\textbf{Theorem: \cite{Gua}} The almost Dirac structure $L$ is
Courant involutive if and only if the exterior derivative $d$
satisfies

\begin{equation}d (C^{\infty}(U)) \subset C^{\infty}(U_{1}),\end{equation}

\noindent i.e. $L$ is involutive if and only if, for any local
trivialization $\rho$ of $U$, there exists a section $X + \xi \in
C^{\infty}((T \oplus T^{*})\otimes \mathbb{C})$ such that

\begin{equation}d \rho = i_{X} \rho + \xi \wedge \rho.\end{equation}

This condition is invariant under rescaling of $\rho$ by a smooth
function.

We can also define the $\bar{\partial}$ operator:

\begin{eqnarray}\bar{\partial} = \pi_{k+1} d: C^{\infty}(U_{k}) \rightarrow
C^{\infty}(U_{k+1}),\\
\partial = \pi_{k-1} d: C^{\infty}(U_{k})
\rightarrow C^{\infty}(U_{k-1}),
\end{eqnarray}

\noindent here $\pi_{k}$ is the projection operator from
$\Lambda^{.}T^{*}$ to $U_{k}$. Then the integrability condition is
equivalent to $d =
\partial + \bar{\partial}$.

\textbf{Definition:} (Twisted almost generalized $G$ structure) We
may twist an almost generalized complex structure by a closed three
form $H$ and the Courant bracket is generalized to:

\begin{equation}[X + \xi, Y + \eta]_{H} = [X, Y] + \mathcal{L}_{X} \eta - \mathcal{L}_{Y}
\xi -\frac{1}{2} d (i_{X} \eta - i_{Y} \xi) + i_{X} i_{Y}
H.\end{equation}

A twisted almost generalized $G$ structure is integrable if and only
if that $L$ is closed with respect to the generalized Courant
bracket. In terms of spinors the integrability condition is then

\begin{equation}d_{H} (C^{\infty}(U)) \subset C^{\infty}(U_{1}),\end{equation}

\noindent where $d_{H} = d + H \wedge$. Similarly one defines
$\bar{\partial}_{H} = \pi_{k+1} d_{H}$ and $\partial_{H} = \pi_{k-1}
d_{H}$. Then the integrability condition of a twisted almost
generalized complex structure is equivalent to $d_{H} =
\partial_{H} + \bar{\partial}_{H}$.

\subsection{String vacua with fluxes turning on}

The supersymmetric transformations for type II theories contain two
ten-dimensional Majorona-Weyl spinor parameters $\epsilon^{1,2}$.
The ten-dimensional manifold is topologically a Minkowski space
times an internal six-dimensional manifold. The ten-dimensional
spinors can be decomposed into spinors in four dimensions times
internal spinors. We are interested in backgrounds preserving
four-dimensional ${\cal N} = 1$ supersymmetry and there is a single
four dimensional conserved spinors. We write

$$\epsilon^{1} = \chi_{+} \otimes \eta^{-}_{1} + \chi_{-} \otimes
\eta^{1}_{-},$$

$$\epsilon^{2} = \chi_{+} \otimes \eta^{-}_{2} + \chi_{-} \otimes
\eta^{2}_{-},$$

for type IIA and

$$\epsilon^{i} = \chi_{+} \otimes \eta^{i}_{+} + \chi_{-}
\otimes \eta^{i}_{-}, i= 1, 2 $$

for type IIB with $\chi_{+}$ any four-dimensional spinor and
$\chi_{-}$ being its Majorana conjugate. We have $(\eta^{i}_{+})^{*}
= \eta^{i}_{-}$ so that $\epsilon^{i}$ are Majorana in ten
dimensions.

Given two spinors we can produce two pure spinors by bispinor
construction:

$$\Phi_{+} = \eta^{1}_{+} \otimes \eta^{+*}_{2}, \Phi_{-} =
\eta^{1}_{+} \otimes \eta^{2*}_{-}.$$

They are compatible pure spinors in the sense that they have exactly
three common annihilators.

If we have one pure spinor we would have structure group reduction
$SU(3,3) \subset SO(6,6)$. When we have tow compatible pure spinors
we would have further structure group reduction $SU(3) \times SU(3)
\subset SO(6,6)$. Thus $\Phi_{\pm}$ defines an almost $SU(3) \times
SU(3)$ structure on $T \oplus T^{*}$.

Preserved supersymmetry imposes differential conditions on the
Clifford(6) spinors. As a consequence, the pure Clifford(6,6)
spinors, as formal sums of differential forms by bispinor
construction, have to obey differential conditions. In order to
preserve ${\cal N} = 1$ supersymmetry, the conditions are
\cite{LMTZ, GMPT1, GMPT2, Witt1}:

$$(d+ H \wedge) \Phi_{1} = 0, (d + H \wedge) \Phi_{2} = * F,$$

where $F = F_{0} + F_{2} + F_{4} + F_{6} + F_{8}$ (for type IIA) or
$F = F_{1} + F_{3} + F_{5} + F_{7} + F_{9}$ (for type IIB) are
Ramond-Ramond fluxes consisting of formal sums of even or odd
degree. Furthermore the fluxes $H, F$ obey Bianchi identities: $dH=
0, d_{H} F = \delta (\cup_{i} D_{i})$, where $D_{i}$ are generalized
submanifolds or orientifolds.

In case of $RR= 0$ we would have ${\cal N} = 2$ string vacua.

There are ${\cal N} = 1$ string vacua with NS fluxes only. We have
one pure spinor and it gives a $SU(3)$ structure with an almost
complex structure and an almost symplectic structure. Those
structures give a (3,0) form $\Omega$ and a (1,1) form $J$.
Preserving supersymmetry implies they satisfy the following
equations \cite{GMW}:

$$(d + H \wedge) (e^{2 \phi} \Omega) = 0, e^{2 \phi} d(e^{-2 \phi} J) =
*H, d(e^{2 \phi} J^{2}) = 0.$$

\section{The moduli space of generalized string vacua}

In this section we shall give a proper definition of the moduli
space of generalized string vacua. Depending on what fluxes are
turning on we shall divide them into several cases.

\subsection{Generalized Calabi-Yau structure}

Let $\Phi = \Sigma_{i=1}^{l} \phi_{i}$ be a sum of forms from a pure
spinor and ${\cal B}(V)$ the orbit of $Cpin(V \oplus V^{*})$ through
$\Phi$. We fix the orbit ${\cal B}(V)$ and go to a manifold $M$. The
orbit ${\cal B}(V)$ yields the orbit in $\oplus_{i=1}^{l}
\Lambda^{*}T_{x}^{*}M$ for each point $x \in M^{n}$. We then have a
fibre bundle ${\cal B}(M)$ by

$${\cal B}(M) := \cup_{x \in M^{n}} {\cal B}(T_{x}M) \rightarrow
M.$$

The set of global sections of ${\cal B}(M)$ is denoted by ${\cal
E}_{\cal B}(M)$. The space of generalized Calabi-Yau structures is:

\begin{equation}\bar{\mathcal{M}} = \{ \Phi \in \mathcal{E}_{\cal B} (M) |
d \Phi = 0 \}.\end{equation}

We need to mode out the action of diffeomorphisms isotopic to
identity and the action of exact two forms. They form a space by
taking a semi-product:

\begin{equation}0 \rightarrow \Omega^{1}(M) \rightarrow \tilde{Diff_{0}(M)}
\rightarrow Diff_{0}(M) \rightarrow 0.\end{equation}

Diffeomorphsims isotopic to identity are generated by vector fields
$X$. In addition to action of diffeomorphisms we also have action of
$B$ fields. We have the action of $X + \xi$ on $\Phi$ as:

$$L_{X + \xi} \Phi = d(i_{X} \Phi + \xi \wedge
\Phi).$$

The universal moduli space of generalized Calabi-Yau structures is
defined as:

\begin{equation}
\tilde{\mathcal{M}} = \bar{\mathcal{M}} / \tilde{Diff_{0}(M^{6})}
\end{equation}

In fact the above universal moduli space is the usual Universal
moduli space for the case of complex structures over a Riemann
surface. We may define the moduli space of generalized Calabi-Yau
structures as:

\begin{equation}
\mathcal{M} = \tilde{\mathcal{M}} / \tilde{Diff(M^{6})}
\end{equation}

\noindent where $\tilde{Diff(M^{6})}$ is the full group of
diffeomorphisms together with a semi product of closed two forms
with integer coefficients.

\subsection{${\cal N}=2$ Type II string vacua}

To incorporate Nervu-Schwarz fluxes we simply twist the exterior
derivative $d$ by $d_{H} = d + H \wedge$. There is a subtlety here
when we try to generalize the Lie derivative with respect to $F = X
+ \xi \in T \oplus T^{*}$. The generalization is:

\begin{equation}
\mathcal{L}_{F} \Phi = F . d_{H} \Phi + d_{H}(F . \Phi)
\end{equation}

The reason to do such a modification is that $H$ changes under a
diffeomorphism. It changes $H$ to $H + d(i_{X} H)$. We knew that $H$
is the field strength of a $B$ field which defines a gerbe
background. $B$ field changes to $B + i_{X} H$ under a
diffeomorphism generated by $X$.

 It is consistent with the fact that a diffeomorphism transforms
integrable structures to integrable structures. For an integrable
twisted generalized $G$ structure defined by $\Phi$ we have $d_{H}
\Phi = 0$, then we have $\mathcal{L}_{F} \Phi = d_{H}(F . \Phi)$. So
under a diffeomorphism generated by $F, \Phi$ stays in the same
$d_{H}$-cohomology class.

The space of $\mathcal{N} = 2$ string vacua is:

\begin{equation}\tilde{\mathcal{M}} = \{ \Phi = (\Phi_{1}, \Phi_{2}) \in \mathcal{E}(M^{6})|
d_{H} \Phi_{1} = 0, d_{H} \Phi_{2} = 0, ||\Phi_{1} = \Phi_{2}|| \} /
\tilde{Diff_{0}(M^{6})}.
\end{equation}

The moduli space of ${\cal N} = 2$ string vacua is:

\begin{equation}
\mathcal{M} = \tilde{\mathcal{M}} / \tilde{Diff(M^{6})}
\end{equation}

\noindent where $\tilde{Diff(M^{6})}$ is the full group of
diffeomorphisms together with a semi product of closed two forms
with integer coefficients.

The above structure is the same as generalized Calabi-Yau metric
structure. A generalized Calabi-Yau metric geometry is defined by a
generalized K$\ddot{\rm a}$hler structure $({\cal J}_{1}, {\cal
J}_{2})$ where each generalized complex structure has
holomorphically trivial canonical bundle, i.e. their canonical line
bundles have non-vanishing closed sections $\rho_{1}, \rho_{2} \in
C^{\infty}(\Lambda^{.} T^{*} \otimes {\bf C})$. We also require that
the lengths of these sections are related by a constant, i.e.
$(\rho_{1}, \bar{\rho_{1}}) = c (\rho_{2}, \bar{\rho_{2}}).$

A generalized Calabi-Yau metric structure may also be twisted by a
three form $H$, by requiring that $({\cal J}_{1}, {\cal J}_{2})$ is
an $H$-twisted generalized K$\ddot{\rm a}$hler structure defined by
$d_{H}$-closed forms $\rho_{1}, \rho_{2}$ satisfying the above
length constraint.

\subsection{${\cal N} = 1$ Type II string vacua}

If we turn on all fluxes the equations for ${\cal N} = 1$ vacua of
generalized structures are:

$$d_{H} \Phi_{1} = 0,$$

$$d_{H}\Phi_{2} = * F.$$

We also have the Bianchi identity $d_{H} F = \delta,$ where $\delta$
is supported over magnetic sources which are usually orientifolds
and generalized submanifolds of the manifold. To be specific, we
consider the case that the $\delta$ function is supported on union
of generalized submanifolds. We formulate the problem as Hodge
systems over a non-compact manifold with divisors union of
generalized complex submanifolds.

We then have the space of $\mathcal{N} = 1$ string vacua as:

\begin{equation}\tilde{\mathcal{M}} = \{(\Phi_{1}, \Phi_{2})
\in \mathcal{E}(M^{6})| d_{H} \Phi_{1} = 0, d_{H} \Phi_{2} =
* F \}/\tilde{Diff_{0}(M^{6})}.
\end{equation}

The moduli space of ${\cal N} = 1$ type II string vacua is:

\begin{equation}
\mathcal{M} = \tilde{\mathcal{M}} / \tilde{Diff(M^{6})}
\end{equation}

\noindent where $\tilde{Diff(M^{6})}$ is the full group of
diffeomorphisms together with a semi product of closed two forms
with integer coefficients.

\section{Deformations of generalized structures}

\subsection{Deformations of generalized Calabi-Yau structures}

We now describe deformations of generalized Calabi-Yau structures by
constructing a complex. Recall that

\begin{equation}CL^{k + 1} = \Lambda^{k}(T \oplus T^{*})/\{u. v + v. u = 2 <u, v>
Id\}.\end{equation}

Let $E^{k}(M) = CL^{k + 1} . \Phi$. Since $\Phi$ is integrable we
have that $d \Phi = 0$. In general this implies that $d E^{k}
\subset E^{k+1}$. We then have a complex:

\begin{equation} ... \rightarrow E^{0}(M) \stackrel{d^{1}}{\rightarrow} E^{1}(M)
\stackrel{d^{2}}{\rightarrow} E^{2}(M) \rightarrow ...
\end{equation}

From \cite{Goto} we knew that $H^{1} = Ker d^{2} / Im d^{1}$ is the
space of infinitesimal deformations of the generalized complex
structures.

We also have a chain map of the above complex into de Rham complex.
Since the forms of pure spinors are closed we have the period map:

\begin{equation} p: H^{1} \rightarrow H^{*}(M).\end{equation}

\textbf{Definition (Elliptic):} An orbit of generalized structures
${\cal B}(V)$ is elliptic if the deformation complex is an elliptic
complex.

\textbf{Definition (Topological structure):} ${\cal B}(V)$ is a
topological structure if the period map $p^{k}_{\cal B}$ is
injective for $k = 1, 2$.

From the following lemma \cite{Goto} we have that if the $d
d^{J}$-lemma is true then the above period map is injective. Hence
the generalized structure is a topological structure.

$d d^{J}-Lemma:$ We say that a manifold have $d d^{J}$ property if
for any exact form $\alpha = d \tau$ which is also $d^{J}$ closed
form then it can be written as $d \tau = d d^{J} \beta.$

\textbf{Lemma:} Suppose that the $d d^{J}$ lemma is true for a
manifold $M$, then for any form $d\tau$ satisfying $d^{J} d \tau =
0,$ there exists $X + \xi$ so that $d \tau = d(i_{X} \Phi + \xi
\wedge \Phi).$

\textbf{Remark:} The construction here is also true for twisted
exterior derivative $d_{H} = d + H \wedge$. The above lemma is true
for a twisted generalized $G$ structure as long as the $d_{H}
d_{H}^{J}$ lemma is true. The proof is in the same way as in
\cite{Goto}.

\textbf{Remark:} A consequence of the above lemma is that the effect
of actions of diffeomorphisms and one forms is the same as taking
cohomology. This would mean that the generalized structure is a
topological structure. This way we get a very simple description of
the moduli space by simply taking cohomologies of the formal sum of
a pure spinor.

\subsection{Deformations of generalized Calabi-Yau metrical
structures}

The equations for ${\cal N} = 2, RR = 0$ string vacua are:

$$d_{H} \Phi_{1} =0, d_{H} \Phi_{2} =0, ||\Phi_{1}|| =
||\Phi_{2}||.$$

Those are called twisted generalized Calabi-Yau metric structures
and they describe all (2,2) nonlinear sigma models. The Universal
moduli space of twisted generalized Calabi-Yau metric structures is:

$$\tilde{\cal M} = \{(\Phi_{1}, \Phi_{2})|d_{H} \Phi_{1} =0, d_{H} \Phi_{2} =0, ||\Phi_{1}|| =
||\Phi_{2}||\} /Diff_{0}(M) \times \Omega^{1}.$$

There is a mapping class group $MCG = Diff(M) /Diff_{0}(M) \times
H^{2}(M; {\bf Z})$ acting on ${\cal M}$. The true moduli space of
${\cal N} = 2$ vacua is ${\cal M} = \tilde{\cal M} /MCG.$

Each pure spinor gives a generalized complex structure. We see that
a generalized Calabi-Yau structure must be a generalized K$\ddot{\rm
a}$hler structure.

\textbf{Definition: (Twisted Generalized K$\ddot{\rm a}$hler
structure)} A twisted generalized K$\ddot{\rm a}$hler structure
consists of two commuting twisted generalized complex structures
${\cal J}_{1}, {\cal J}_{2}$ such that ${\cal G} = - {\cal J}_{1}
{\cal J}_{2}$ is positive definite over $T \oplus T^{*}$.

Since ${\cal G} ^{2} = I, T \oplus T^{*} = C_{+} \oplus C_{-}$,
where $C_{\pm} = \{ {\cal G} = \pm 1 \}$. We have that $<,
>|_{C_{+}} > 0, <, >|C_{-} < 0, <, >|C_{+}$ gives a metric over
$C_{+}$.

Let $* = a_{1} ... a_{n}$ be the product in $CL(C_{+}) , CL(T \oplus
T^{*})$ of an oriented orthonormal basis for $C_{+}$. This volume
element acts on the differential forms via the spin representation.
When $b= 0$, it gives the Hodge star operator $*: \Lambda^{.}T^{*}
\rightarrow \Lambda^{.}T^{*}$. In general we have $*^{2} =
(-1)^{n(n-1)/2}$. Let $\alpha$ be a $k$ form, define $\sigma \alpha
= (-1)^{k (k-1)/2} \alpha$. Over $\Lambda^{*}T^{*}$ we have an inner
product:

$$h(\alpha, \beta) = \int_{M} <\alpha, \sigma(*) \bar{\beta}>,$$

\noindent which we call the Born-Infeld inner product. We will
calculate the adjoint of the twisted exterior derivative $d_{H}$
with respect to this inner product.

The $d_{H} d_{H}^{J}$ lemma for a twisted generalized K$\ddot{\rm
a}$hler manifold is established in \cite{Gua-Hodge}. We reproduce it
here for convenience. This will be crucial for us to imbed the
moduli space into the space of de Rham cohomologies.

For each twisted generalized complex structure ${\cal J}$ we have
decomposition of forms into

$$U = U_{0} < U_{2} < ... < U_{2n} = \Lambda^{ev/od} T^{*} \otimes
{\bf C},$$

$$L^{*} U_{0} = U_{1} < U_{3} < ... < U_{2n-1} = \Lambda^{od/ev}
T^{*} \otimes {\bf C}$$

\noindent where $U_{k}$ is an invariant eigenspace of ${\cal J}_{1}$
with eigenvalue $-ik$ acting on forms through spin representation. A
generalized complex structure ${\cal J}$ preserves the canonical
pairing between tangent and cotangent spaces. It is then an element
of $so(n,n)$. It then acts on the space of forms $\Lambda^{*}T^{*}$
by spin representation.

With respect to the second twisted generalized complex structure
which commutes with the first twisted generalized complex structure
we have further decomposition:

\begin{equation}U_{k} = U_{k, |k|-n} \oplus U_{k, |k| -n +2} \oplus ... \oplus
U_{k, n -|k|}, U_{p, q} = U^{1}_{p} \cap U^{2}_{q}.\end{equation}

It gives the Hodge decomposition. For a twisted generalized
K$\ddot{\rm a}$hler manifold, we have that

\begin{eqnarray}d_{H} = \delta_{+} + \delta_{-} + \bar{\delta}_{+} +
\bar{\delta}_{-},\\
\bar{\partial}_{1} = \bar{\delta}_{+} + \bar{\delta}_{-},
\bar{\partial}_{2} = \bar{\delta}_{+}+ \delta_{-}.\end{eqnarray}

Gualtieri announced the following generalization of K$\ddot{\rm
a}$hler identities \cite{Gua-Hodge}:

\begin{equation}
\bar{\delta}_{+}^{*} = - \delta_{+}, \bar{\delta}_{-}^{*} =
\delta_{-}.
\end{equation}

Here $\bar{\delta}^{*}_{+} = * \delta_{+}
*^{-1}$ is the adjoint. Those identities would imply the $d_{H} d_{H}^{J}-lemma$.

From \cite{Goto}, a generalized Calabi-Yau metric structure can be
described by two generalized $Sl(n, {\bf C})$ structures. An $Sl(n,
{\bf C})$ structure is a complex form of type $(n, 0), \Omega_{V}$
with respect to a complex structure $J$.

Conversely for each $Sl(n, {\bf C})$ structure $\Omega_{V}$ we
define a complex subspace $ker \Omega_{V}$ by $ker \Omega_{V} = \{v
\in V_{\bf C} | i_{v} \Omega_{V} = 0\}$. We have then $V_{\bf C} =
\ker_{\Omega_{V}} \oplus \bar{\ker \Omega_{V}}$. This determines a
complex structure $J$.  In other words an $Sl(n, {\bf C})$ structure
$\Omega_{V}$ gives a complex structure $J$ on $V$.

Let $\Omega_{V}$ be an $Sl(n, {\bf C})$ structure and $\omega_{V}$ a
real two form on $V$. We define a bilinear form $g$ by $g(u,v) =
\omega(Ju, v)$. A pair $(\Omega_{V}, \omega_{V})$ is a Calabi-Yau
structure on $V$ if the followings hold:

$1) \Omega_{V} \wedge \omega_{V} = 0, $

$2) \Omega \wedge \bar{\Omega} = c_{n} \omega^{n},$

3) The corresponding bi-linear form $g$ is positive-definite.

Given Gualtieri's $d_{H} d_{H}^{J}$ lemma and by slightly modifying
the proof of Goto we have:

\textbf{Theorem:} The twisted generalized Calabi-Yau orbit is an
elliptic and topological orbit. This would imply that the moduli
space can be embedded into the space of twisted cohomologies
locally.

\subsection{Deformations of generalized structures of ${\cal N} = 1$ vacua}

The equations for ${\cal N} = 1$ vacua of generalized structures
are:

$$d_{H} \Phi_{1} = 0, $$

$$d_{H} \Phi_{2} = * F.$$

We also have the Bianchi identity $d_{H} F = \delta,$ where $\delta$
is supported over magnetic sources \cite{VZ}. To be specific, we
consider the case that the $\delta$ function is supported on union
of generalized submanifolds with respect to the first generalized
complex structure. We formulate the equations as Hodge systems over
a non-compact manifold with divisors union of generalized complex
submanifolds, i.e. $M = \bar{M} - \cup_{i} D_{i}$. We further assume
that $D_{i}$ are divisors with normal crossing, i.e. locally $D_{i}$
is described by equations $z_{1} ... z_{r} = 0$ in the case of
complex structure with respect to complex coordinates, or by $p_{1}
= c_{1}, ..., p_{r} = c_{r}$ in the case of symplectic structure
with respect to a polarization. By a generalized Darboux theorem, a
generalized complex structure is locally a product of complex and
symplectic structures.

Over $M$ we need to choose a proper functional space to work with.
It turns out that the proper functional space consists of forms
which are square integrable and their exterior derivatives are
square integrable \cite{de Rham}.

Over this functional space the Hodge systems are soluble. The key
step is to construct a Green current which gives fundamental
solutions of the Hodge systems. This can be done over this
functional space. Let $d_{H}^{*}$ be the adjoint of $d_{H}$. We then
have Laplacian $\Delta_{H} = d_{H} d_{H}^{*} + d_{H}^{*} d_{H}$. The
Green current is a solution of the equation $\Delta_{H} G =
\delta(\cup_{i} D_{i})$. The Green current satisfies equivalent
equations: $d_{H} G = 0, d_{H} *G = *F.$ We have $d_{H} (\Phi_{2} -
*G) = 0$ so that $\Phi_{2} - * G$ represents an integrable
generalized complex structure ${\cal J}_{2}^{'}$. We see that the
Green current represents a co-chain of $d_{H}$-cohomology which
represent the union of divisors.

Alternatively, the Green current can be constructed by solving the
Dirac equation $D G = F$, where $D = d_{H} + d_{H}^{*}$. We see then
$D(\Phi_{2} - * G) = 0$.

${\cal J}_{1}$ and ${\cal J}_{2}^{'}$ commutes. This follows from
that the divisors are generalized submanifolds with respect to
${\cal J}_{1}$. Hence the divisors are invariant with respect to
${\cal J}_{1}$. The Green current is then invariant with respect to
${\cal J}_{1}$.

For the first generalized complex structure coming from $\Phi_{1}$,
we have decomposition of forms as eigenspaces of the action by
${\cal J}_{1}$ in the spin representation. For the second
generalized complex structure coming from $\Phi_{2}$ it is not
integrable. We can replace it by $\Phi_{2} - * G$ which is a new
integrable generalized complex structure ${\cal J}_{2}^{'}$. We have
decomposition of forms as eigenspaces of the action by ${\cal
J}_{2}^{'}$ in the spin representation. ${\cal J}_{1}$ and ${\cal
J}_{2}^{'}$ commutes. So we have Hodge decomposition over $M$ by
using generalized complex structures ${\cal J}_{1}$ and ${\cal
J}_{2}^{'}$.

We may decompose the twisted exterior operator $d_{H}$ into four
operators $\delta_{+} + \delta_{-} + \bar{\delta}_{+} +
\bar{\delta}_{-}$ with $\bar{\partial}_{1} = \bar{\delta}_{+} +
\bar{\delta}_{-}$ and $\bar{\partial}_{2} = \bar{\delta}_{+} +
\delta_{-}$. Here the decomposition is respect to ${\cal J}_{1}$ and
${\cal J}_{2}^{'}$.

 We still have the generalized K$\ddot{\rm a}$hler identities:
$\delta^{*}_{+}= - \delta_{+}, \bar{\delta}_{-}^{*} = \delta_{-}.$
 This would imply the $\partial_{H} \bar{\partial}_{H}$ lemma. Proof of the
generalized K$\ddot{\rm a}$hler identities is to compute adjoint of
the $\delta_{+}$ with respect to the Born-Infeld metric. We can
perform similar computations as for K$\ddot{\rm a}$hler manifolds.
The conditions of square integrable of forms and their exterior
derivatives make all the computations possible.

Hence the period map is injective so the universal  moduli space of
${\cal N} = 1$ generalized vacua can be imbedded into the space of
cohomologies. We have two pure spinors. For $\Phi_{1}$ we have
$d_{H} \Phi_{1} =0$ and the moduli would be just the
$d_{H}$-cohomology classes of $\Phi_{1}$. For $\Phi_{2}$ we have
$d_{H} \Phi_{2} = * F$ and the moduli would be the
$d_{H}$-cohomology classes of the co-chain represented by
$\Phi_{2}$. All this follows from the $\partial_{H}
\bar{\partial}_{H}$ lemma.

There is again a mapping class group MCG acting on ${\cal M}$. Here
we need to modify the definition of $H^{2}(M,{\bf Z})$ to
$H^{2}_{c}(M, {\bf Z})$ of cohomology with compact support. And
$Diff_{c}(M)$ consists of diffeomorphisms leaving divisors
invariant.

Finally we have the true moduli space of ${\cal N} = 1$ generalized
vacua. MCG induces an action on the space of cohomologies which
gives a local system. The flat connection of such a local system is
the usual Gauss-Manin connection. The usual Picard-Fuchs equations
may follow from this and thus give a basis of topological B model.

\section{Special geometry of the moduli space of generalized string vacua}

Over the moduli space we have a special geometry which means that we
have a flat symplectic structure, an integrable complex structure
and a K$\ddot{\rm a}$hler metric. In this section we shall construct
such a special geometry (see also \cite{Hitchin}, \cite{GLW1,
GLW2}).

\subsection{The symplectic structure over $\mathcal{M}$}

We define a symplectic structure over $\mathcal{M}:$

\begin{equation}
\omega (\Phi_{1}, \Phi_{2}) = \int_{M^{6}} <\Phi_{1}, \Phi_{2}>.
\end{equation}

Here we have the Mukai Pairing:

\begin{equation}
<\Phi_{1}, \Phi_{2}> = \Sigma_{p} (-1)^{\left\lfloor
\frac{p+1}{2}\right\rfloor} \Phi_{1,p} \wedge \Phi_{2,
n-p}.
\end{equation}

To show that it defines a symplectic structure over the moduli space
we may check that the integral depends on $d_{H}$-cohomology classes
only. This follows from a formular proved by Hitchin:

\begin{equation}\int_{M} \sigma(\hat{\alpha}) \wedge d_{H} \beta =
- \int_{M} \sigma(d_{H} \hat{\alpha}) \wedge \beta.\end{equation}

Since we are integrating it over the manifold it is also
diffeomorphism invariant. Actually the Clifford action of the
tangent bundle and the cotangent bundle stays in the same $d_{H}$
cohomology classes.

Since the definition only depends on integrating forms and not
depends on anything else it is a constant form so we have $d \omega
= 0.$ According to Darboux's theorem around each point we can take a
special Darboux coordinates so that

\begin{equation}\omega = \Sigma dx^{K} \wedge dy_{K}.\end{equation}

Those coordinates are very useful in constructing special
geometries.

\subsection{The complex structure over ${\mathcal M}$}

We knew that stable spinors can be decomposed into sum of pure
spinors:

\begin{equation}\Phi = \rho + \hat{\rho}.\end{equation}

We have an involution map:

\begin{equation}X: \rho = \phi + \bar{\phi} \rightarrow \hat{\rho}= -i\phi + i
\bar{\phi},
\end{equation}

The differential of $X$ defines an almost complex structure over
stable spinors because $DX . DX = -Id.$ Actually $J = DX$ defines an
integrable complex structure over $\mathcal{M}$. We can prove this
by using Darboux coordinates. Consider the dual basis of $dx^{K},
dy_{K}$ of the tangent space we can express the symplectic form as:

\begin{equation}\omega(\Phi, \bar{\Phi}) = \Sigma_{K} (\bar{Z}^{K} F_{K} - Z^{K}
\bar{F}_{K}).\end{equation}

One can show that $Z^{K}, F_{K}$ are two independent complex
coordinates.

\subsection{Hitchin functional and the K$\ddot{\rm a}$hler metric over $\mathcal{M}$}

By using the symplectic structure $\omega$ and the complex structure
$J$ we may define a symmetric two tensor $g(., .) = \omega(., J.)$.
We shall see that this gives rise a metric. This metric is actually
K$\ddot{\rm a}$hler and its K$\ddot{\rm a}$hler potential is the
Hitchin functional.

Recall that the generalized Hitchin functional is:

\begin{equation}H(\phi, F) = \int_{M^{6}} -i <\phi, \hat{\phi}> + <F, \eta>,
\end{equation}

\noindent where $\phi = \phi_{0} + d_{H} \eta$ is a stable spinor
and $F$ is a formal sum of forms which represent the Ramond-Ramond
fluxes. It is a function over $\mathcal{M}$ because it depends only
on the $d_{H}$ cohomology class and it is a diffeomorphic invariant.

From the definition of the Hitchin functional we see that
$\hat{\rho} = \delta H / \delta \rho.$ Since the derivative of the
map $X: \rho \rightarrow \hat{\rho}$ gives rise to the complex
structure $J$ we have that the metric $\omega(J., .)$ is the same as
the one using the Hitchin functional:

\begin{equation} g_{\alpha \beta} = \partial^{2} H /\partial
\chi^{\alpha} \partial \bar{\chi}^{\beta}
\end{equation}

Here $\partial \chi_{\alpha}$ is a basis of the holomorphic tangent
space of the moduli space. This metric is actually K$\ddot{\rm
a}$hler and the K$\ddot{\rm a}$hler potential of the metric is: $K =
- \log H.$

We finally have:

\begin{equation}e^{-K(\Phi)} = H(\Phi) = i \omega(\Phi, \bar{\Phi}) =
\Sigma_{K} i(\bar{Z}^{K} F_{K} - Z^{K} \bar{F}_{K}).
\end{equation}

This generalize a formular for the moduli space of complex
structures of Calabi-Yau manifold:

\begin{equation}
K^{2,1} = - \log(i \int \Omega \wedge \bar{\Omega}),\\
K^{1,1} = - \log(i \int_{M} \omega \wedge \omega \wedge \omega).
\end{equation}

\section{Examples}

    Here are a few examples of generalized Calabi-Yau structures and
    twisted generalized K$\ddot{\rm a}$hler structures. There are more examples in
    the literature.\\

    1) Nilmanifolds\\

    A nilmanifold is a homogeneous space $M = G/ \Gamma$, where $G$
    is a simply-connected nilpotent real Lie group and $\Gamma$ is a
    lattice of maximal rank in $G$. The nilmanifold can be described
    by giving the differentials of a set $\{e_{1}, e_{2}, ...,
    e_{6}\}$ of linearly independent left-invariant 1-forms. In the
    nilmanifold literature one uses the array $(0,0,0,12,13,14+35)$
    to describe a nilmanifold with de Rham complex generated by
    1-forms $\{e_{1}, e_{2}, ..., e_{6}\}$ and such that
    $de_{1} = de_{2} = de_{3} = 0,$ while $de_{4} = e_{1} \wedge
    e_{2}, de_{5} = e_{1} \wedge e_{3},$ and $de_{6} = e_{1} \wedge
    e_{4} + e_{3} \wedge e_{5}$.
    For six dimensional nilmanifolds there are 5 classes of
    nilmanifold which admit no known complex or symplectic
    structure. They are:

    $(0,0,12,13,14+23, 34+52);$

    $(0,0,12,13,14,34+52);$

    $(0,0,0,12,13,14+35);$

    $(0,0,0,12,23,14+35);$

    $(0,0,0,0,12,15+34).$

    Those five families admit generalized complex structures. In
    each case, the canonical bundle is holomorphically trivial.
    Hence they are examples of generalized Calabi-Yau structures.
    For more details, see \cite{Gua}, \cite{CG}. \\

    2) Homogeneous space \\

    Any compact even-dimensional Lie group admits left- and
    right-invariant complex structure $J_{L}, J_{R}$, and that if
    the group is semi-simple, they can be chosen to be Hermitian
    with respect to the bi-invariant metric induced from the Killing
    form $<,>$. We would have a 3-form $H, H(X,Y,Z) = <[X,Y],Z>$.
    It turns out that $(<,>, J_{L}, J_{R})$ forms an H-twisted
    generalized K$\ddot{\rm a}$hler structure, see \cite{Gua}. \\

    3) Connected sum of $S^{3} \times S^{3}$\\

    $S^{3} \times S^{3}$ is a manifold with two complex
    structures. However they are not a K$\ddot{\rm a}$hler manifold since
    $h^{1,1} = 0$. Consider $S^{3}$ as a copy of $SU(2)$. The Lie algebra of
    $su(2)$ are Pauli matrices $\sigma_{1}, \sigma_{2}, \sigma_{3}$.
    We have then two complex structures. One of them is given by:

    $\omega = e^{i \pi /4} (\sigma^{1} + i \sigma^{2}) \wedge
    (\hat{\sigma}^{1} - i \hat{\sigma}^{2}) \wedge(\sigma^{3} + i
    \hat{\sigma}^{3}),$

    $\Omega = \sigma^{1} \wedge \sigma^{2} - \hat{\sigma}^{1}
    \wedge \hat{\sigma}^{2} + \sigma^{3} \wedge \hat{\sigma}^{3}.$

    The other complex structure is:

    $\omega = e^{i \pi /4} (\sigma^{1} + i \hat{\sigma}^{1}) \wedge
    ({\sigma}^{2} + i \hat{\sigma}^{2}) \wedge(\sigma^{3} + i
    \hat{\sigma}^{3}),$

    $\Omega = \sigma^{1} \wedge \hat{\sigma}^{1} + \sigma^{2}
    \wedge \hat{\sigma}^{2} + \sigma^{3} \wedge \hat{\sigma}^{3}.$

    We may take connected sums of $S^{3} \times S^{3}$. It is found that
    manifolds of $k \ge 2$ copies of $S^{3} \times S^{3}$ satisfy $\partial
    \bar{\partial}$ Lemma property (\cite{GIP}). Such manifolds appears in the
    smoothing of Calabi-Yau manifolds.

\section{Conclusions and prospects}

    The importance of flux compactifications have been realized for
    a variety of problems in string theory and in its applications to
    field theory and cosmology. Systematic studies are carried out
    in recent years. New geometric structures such as generalized G
    structures arises and it fits naturally to describe string vacua
    with fluxes turning on.

    In string compactifications we need to know deformations of a
    given string vacua and the special geometry they obey. As a
    first step we need to identify the moduli space of vacua and
    its tangent space. It turns out there are several cases
    depending on what fluxes are turning on. They are respectively
    generalized Calabi-Yau, generalized Calabi-Yau metric structures
    and the most general case of ${\cal N} = 1$ generalized vacua with both
    NN fluxes and RR fluxes turning on. In this paper we give
    definitions of moduli spaces for those cases respectively and
    identify their tangent spaces.

    It turns out that the crucial step is an elementary lemma called
    the $\partial \bar{\partial}$ lemma which implies the $d d^{J}$
    lemma. Those lemmas play key roles to show that the period map
    is injective so that the moduli space can be imbedded into the
    space of de Rham cohomologies.

    As soon as one identifies the space of vacua one can establish a
    special geometry on the moduli space of vacua. One confirms that
    the Hitchin functional appears naturally as the potential of
    the K$\ddot{\rm a}$hler metric over the moduli space.

    There are a number of directions one can follow naturally.
    There are works to identify the open string moduli
    \cite{KM1, KM2, KM3, KT, EM}. It would be interesting to consider
    the full moduli space and to identify superpotential.

    It is interesting to study dualities among different kinds of
    compactifications. In heterotic string compactifications the equations
    of supersymmetric solutions were derived by Strominger \cite{Strominger}.
    There are much works recently in finding new solutions and applications
    to heterortic string compactifications \cite{FY, BBFTY, BTY1, GY}. It would be
    interesting to study dualities of flux vacua of heterotic strings with
    flux vacua of type II strings. It is also interesting to study
    supersymmetric solutions in M theory \cite{Tsi, Hu}.

    There is an string duality of vacua between type IIA and type IIB
    called mirror symmetry. There are a number of interesting
    proposals for generalized geometry that the mirror
    symmetry is simply interchanging two generalized complex
    structures \cite{AFT, LRRUZ}. Since mirror symmetry acts on topological
    models one needs to study such models for the generalized geometries.
    It would be interesting to extend works of \cite{Witten, BCOV}
    for the setting of generalized geometry.

    Our study can be considered as a study of variations of Hodge
    structures. It then raises many problems to extend the work on
    variations of Hodge structures over K$\ddot{\rm a}$hler manifolds to
    generalized K$\ddot{\rm a}$hler manifolds. Since we have the crucial $d
    d^{J}$ lemma we would expect many works extends to the more
    general case. See \cite{CV, JYZ} for works on variations of Hodge structures.

    The study of flux compactifications are limited largely due to
    the lack of examples. There are quite a few examples appeared in
    recent years. It is still lacking a theorem like Calabi-Yau
    theorem for the category of generalized complex geometry. For
    this even a proper generalization of Calabi's conjecture would
    be very interesting.

    There are many applications in physics. For example, there are
    some works on generalized black hole attractor mechanisms, works on
    gauge-gravity correspondence and works on moduli stabilization
    especially for inflation models. We expect that the work on
    flux compactifications would help substantially on all those
    problems.\\

\textbf{Acknowledgements:} We wish to thank M. Atiyah, D. Gross, N.
Hitchin, K. Liu, A. Strominger, H. Tye, C. Vafa, E. Witten, Y. S. Wu
and S. T. Yau for their comments and encouragements in various
occasions. Discussions with R. Goto, M. Gualtieri, C. Leung, L.
Martucci, A. Todorov, A. Tomassielo, Yang Jie, Yin Zheng and Zhou
Junjie are also very helpful. Part of the work is supported by a
NSFC grant (No. 10771203) and a Chuangxin grant from the Chinese
Academy of Sciences.

\appendix

\section{Mukai pairing and the symplectic structure}

{\defn{Mukai pairing}\\ Let $x\mapsto x^T$ be the main
antiautomorphism of the Clifford algebra $CL(V \oplus V^*)$, i.e.
that determines by the tensor map $v_1\otimes v_2\otimes...\otimes
v_k\mapsto v_k\otimes v_{k-1}\otimes...\otimes v_1$. The Mukai
pairing is defined to be
\begin{equation}(\cdot,\cdot)_M: \wedge^\bullet V^*\otimes\wedge^\bullet V^*\rightarrow\ det\ V^*,\\
(s,t)_M=[s^T\wedge t]_m,\ m=dim\ V.
\end{equation}}

Mukai pairing has the following properties:
\begin{equation}(x\cdot s,t)_M=(s,x^T\cdot t)_M,\ \forall x\in CL(V\oplus V^*).\end{equation}

{\prop The Mukai pairing is invariant under the identity component
of Spin:
\begin{equation}(g\cdot s, g\cdot t)_M=(s,t)_M, \forall g\in Spin_0(V\oplus V^*).\end{equation}
 Therefore the Mukai pairing defines a $Spin_0$ invariant
bilinear form on $S=\wedge^\bullet V^*\otimes (det\ V)^{1/2}$.}

When $m=6$, the Mukai pairing is skew symmetric. We shall begin to
study the algebra by working over the complex numbers, using
$Spin(12,\mathbb{C})$ instead of $Spin(6,6)$ and a complex
six-dimensional vector space $V$ . In this dimension the bilinear
form on each of the 32-dimensional spin spaces $S^{\pm}$ is skew
symmetric, and so these are symplectic representations. We shall fix
$S=S^{\pm}$.

A symplectic action of a Lie group $G$ on a vector space S defines a
moment map

\begin{equation}\overline{\mu}: S\rightarrow
g^*\otimes\wedge^6V^* \end{equation} given by

\begin{equation}\overline{\mu}(\rho)=\frac{1}{2}(\sigma(a)\rho, \rho)_M,\ \forall a\in g,\ \rho\in S.
\end{equation}

Here $G=Spin(V\oplus V^*),\ g=spin(V\oplus V^*)\cong
so(12,\mathbb{C}),\ \sigma: g\rightarrow End(S)$ is the
representation of Lie algebras and $a\in g$. And we identify $g$
with $g^*$ by the inner product \begin{equation}(\cdot,\cdot):
g\otimes g\rightarrow \mathbb{C},\ (X,Y)\rightarrow
tr(XY).\end{equation}

{\defn Let $\overline{\mu}$ be the moment map for the spin
representation $S$ of $Spin(12,\mathbb{C})$, then

\begin{equation}\overline{q}(\rho)=tr\overline{\mu}(\rho)^2
\end{equation}

\noindent is an invariant quartic function on $S$.}

This quartic has a close relationship with pure spinors:

{\prop For $\rho\in S, \overline{\rho}\neq 0$ if and only if
$\rho=\alpha+\beta$, where $\alpha, \beta$ are pure spinors and
$(\alpha,\beta)_M\neq 0$. The spinors $\alpha, \beta$ are unique up
to ordering. }

It is proved by Hitchin \cite{Hitchin}. From the proof, we have
\begin{equation}\overline{q}(\alpha+\beta)=3(\alpha,\beta)^2_M,\\
\overline{\mu}^2=\frac{1}{48}\overline{q}(\rho)I.
\end{equation}

Let us consider a real $6-dim$ vector space $W,\ S=\wedge^{ev}W^*$.
$q(\rho)\neq 0$ implies there are two possibilities: $\alpha$ and
$\beta$ are both real, or $\beta=\bar{\alpha}$, denoted by
$\overline{q}(\rho)>0,\ \overline{q}<0$. (Note: Let $L$ is a real
one-dimensional vector space, $u\in L\otimes L$,  we say $ u>0$ if
$u=s\otimes s$ for some $s\in L$; $u<0$ if $-u>0$.)

Let us fix an orientation $\epsilon\in\wedge^6W^*$ on $W$, define a
symplectic form on $\wedge^\bullet W^*$:
 \begin{equation}\omega: \ \wedge^\bullet W^*\times \wedge^\bullet
 W^*\rightarrow\mathbb{R},\end{equation}
\noindent such that $\omega(\rho_1,
\rho_2)\epsilon=(\rho_1,\rho_2)_M $.

Re-define the moment map $\mu: S\rightarrow g^*$ by
$\rho\mapsto\mu(\rho)(a)=\frac{1}{2}\omega(\sigma(a)\rho,\rho)$. And
let $q(\rho)=tr\mu(\rho)^2$. It is easy to see that
$\overline{\mu}=\mu\otimes\epsilon,\
\overline{q}(\rho)=q(\rho)\epsilon^2$.

Consider the open set
\begin{equation}U=\{\rho\in S:
q(\rho)<0\}
\end{equation}
\noindent acted on transitively by the real group
$\mathbb{R}^*\times Spin(6,6)$.

{\defn Define a homogeneous function of degree 2 on U:
$\phi(\rho)=\sqrt{-q(\rho)/3}$.}

Note from the above proposition we can write
$\rho=\varphi+\bar{\varphi}$ for a pure spinor $\varphi$ such that
\begin{equation}i\phi(\rho)=\omega(\phi,\bar{\phi}).\end{equation}

{\prop  Let $X$ be the Hamiltonian vector field on $U$ defined by
the function $\phi$ using the constant symplectic form on $U\subset
S$. Describe the vector field on the open set $U$ in the
vector space $S$ as a function $X: U\rightarrow S$. Then \\
$\star\ X(\rho)=\hat{\rho}$ where $\rho+i\hat{\rho}=2\varphi;$\\
$\star\ X$
generates the circle action $\varphi\mapsto e^{-i\theta}\varphi$.\\
$\star$ the derivative $DX: U\rightarrow End(S)$ defines an
integrable almost complex structure $J$ on $U$.}

By the definition of Hamiltonian vector field, we have the
derivative at $\rho$ of $\phi$ is a linear map can be written as
\begin{equation}D\phi(\dot{\rho})=\omega(\hat{\rho},\dot{\rho});\end{equation}
The second derivative
\begin{equation}D^2\phi(\hat{\rho}_1,\hat{\rho}_2)=\omega(DX\hat{\rho}_1,
\hat{\rho}_2)=\omega(J\hat{\rho}_1,\hat{\rho}_2)\end{equation}

Suppose $M$ is a compact oriented $6-$manifold with volume form
$\epsilon$, and $\rho$ is a sum of forms, either odd or even, which
lies at each point of $M$ in the open subset $U$ described above.
Such a form is called \emph{stable}. We can then define a volume
functional
\begin{equation}V(\rho)=\int_M\phi(\rho)\epsilon.\end{equation}

Let $H$ be a closed 3-form on $M$, define an operator on forms:
\begin{equation}d_H\alpha=d\alpha+H\wedge\alpha.\end{equation}
It is easy to see that $d_H^2=0$. We can define $d_H-$cohomology.
Consider the variational problem in a fixed $d_H-$cohomology class.
We have

{\thm A $d_H-$closed stable form $\rho\in \wedge^{ev/od}(M)$ is a
critical point of $V(\rho)$ in its $d_H-$cohomology class if and
only if $d_H(\hat{\rho})=0$.}

\emph{Proof}: by computation, we have $\int_M(\hat{\rho}, d_H\alpha
)_M=-\int_M(d_H\hat{\rho}, \alpha)_M.$

At a critical point of $V$, the Hessian $H$ is
\begin{equation}H(d_H\alpha_1, d_H\alpha_2)=-\int_M(d_HJd_H\alpha_1,\alpha_2)_M.
\end{equation}

\section{Hitchin's functional and the variational principle}

Given a stable spinor $\rho$ we define Hitchin's functional as:

\begin{equation}H(\rho) = \int_{M} (\sigma(\rho) \wedge J \rho)_{top}.\end{equation}

We then consider an variational problem of $H(\rho)$ with a fixed
$d_{H}$-cohomology with $d_{H} = d + H \wedge$. The Euler-Lagrange
equations are:

\begin{equation}d_{H} \rho = 0, d_{H} (J \rho) = 0.\end{equation}

From \cite{Witt1} those equations are precisely the integrability
conditions of a twisted almost generalized $G$ structure.

To incorporate the Ramond-Ramond fluxes we generalize Hitchin's
functional as:

\begin{equation}H(\rho) = \int_{M} (\sigma(\rho) \wedge J \rho)_{top} + (\eta \wedge F)_{top},\end{equation}

\noindent where $\rho = \rho_{0} + d_{H} \eta, d_{H} \rho_{0} =0,
d_{H} (J \rho_{0}) = 0$.

The Euler-Lagrange equations of $H(\rho)$ for a fixed
$d_{H}$-cohomology class are \cite{JW}:

\begin{equation}d_{H} \rho = 0, d_{H} (J \rho) = * F.\end{equation}

Those equations are precisely the supersymmetry equations with all
fluxes turning on.

\end{document}